\documentstyle[aps,prl,multicol,floats,epsf]{revtex}

\begin{document}
\draft
\wideabs{

\title{The Hall effect in Zn-doped YBa$_{2}$Cu$_{3}$O$_{7-\delta}$ 
revisited: Hall angle and the pseudogap}

\author{Yasushi Abe$^{1,2}$, Kouji Segawa$^{1}$, and Yoichi Ando$^{1,2}$}

\address{$^1$Central Research Institute of Electric Power
Industry, Komae, Tokyo 201-8511, Japan}
\address{$^{\rm 2}$ Department of Physics, Science University of Tokyo, 
Shinjuku-ku, Tokyo 162-8601, Japan}

\date{Hall-6.tex}
\maketitle

\begin{abstract}
The temperature dependence of the Hall coefficient is measured with a 
high accuracy in a series of 
YBa$_{2}$(Cu$_{1-z}$Zn$_z$)$_3$O$_{6.78}$ crystals 
with 0$\le$$z$$\le$0.013.
We found that the cotangent of the Hall angle, $\cot \theta_H$, 
starts to deviate upwardly from the $T^2$ dependence 
below $T_0$ ($\sim$130 K), regardless of the Zn concentration.  
We discuss that this deviation is caused by the pseudogap;
the direction of the deviation and its insensitivity to the Zn doping
suggest that the pseudogap affects 
$\cot \theta_H$ through a change in the effective mass, rather than
through a change in the Hall scattering rate.
\end{abstract}

}
\narrowtext

The strong temperature dependence of the Hall coefficient $R_H$ of 
the high-$T_c$ cuprates has been considered to be one of the most peculiar 
properties of their unusual normal state \cite{Anderson1}.
The rather complex behavior \cite{Cooper} of $R_H(T)$ 
can be turned into a simpler one 
by looking at the cotangent of the Hall angle \cite{Chien}, 
$\cot \theta_H \equiv \rho_{xx}$/$\rho_{xy}$; 
it has been shown that $\cot \theta_H$ of
cuprates behaves approximately as $T^2$,
regardless of material \cite{Yakabe} and carrier concentration 
\cite{Cooper}.
This remarkable simplicity in the behavior of $\cot \theta_H$
led to the idea \cite{Chien,Anderson2} that $\cot \theta_H$ reflects 
a Hall scattering rate 
$\tau_H^{-1}$, which is different from the scattering rate 
$\tau_{tr}^{-1}$ governing the diagonal resistivity $\rho_{xx}$.
There are two physical pictures to account for this apparent separation 
of the scattering rates: 
One picture considers that two distinct scattering times 
$\tau_{tr}$ and $\tau_H$, possibly associated with different particles,
govern different kinds of scattering events \cite{Anderson2,Coleman}.  
The other picture considers that the scattering time is 
strongly dependent on the position on the Fermi surface (FS) 
and that $\rho_{xx}$ and $\cot \theta_H$ are governed by the 
scattering events on different parts of the FS \cite{Cooper,Hot/Cold}.  

Separately from the above development, 
it has become a common understanding \cite{Batlogg,Ong} 
that in underdoped cuprates 
a pseudogap in the density of low-energy excitations is
developed at a temperature much higher than the 
superconducting transition temperature $T_c$.
In underdoped YBCO, the in-plane resistivity $\rho_{ab}$ 
shows a clear downward deviation from 
the $T$-linear behavior below a temperature $T^*$, which has been
discussed to mark the onset of the pseudogap \cite{Ito}.
This $T^*$ is notably higher than the other characteristic temperature 
$T_g$ determined from the onset of a suppression in the Cu NMR relaxation 
rate \cite{NMR}, which has also been associated with the pseudogap.
The presence of two different temperature scales, $T^*$ and $T_g$, is 
intriguing.  
It was proposed recently that at the upper temperature scale $T^*$ 
the CuO$_2$ plane starts to develop local antiferromagnetic correlations 
\cite{Batlogg} or charged stripe correlations \cite{Emery};
the lower temperature scale $T_g$ corresponds to the 
opening of a more robust pseudogap in the density of states 
\cite{Batlogg},
which can be observed by the angle-resolved photoemission 
\cite{ARPES} or by the tunneling spectroscopy 
\cite{Tunneling}.

It was previously discussed \cite{Ito} that the pseudogap causes 
a deviation from the $T^{-1}$ behavior in $R_H(T)$ at $T^*$.
The conspiring changes in $\rho_{ab}(T)$ and $R_H(T)$ at $T^*$ leave
the $T^2$ behavior of $\cot \theta_H$ unchanged at $T^*$, 
which led to the belief 
that $\cot \theta_H$ is rather insensitive to the opening of the pseudogap.
However, given the recent understanding that the pseudogap has 
two characteristic temperatures $T^*$ and $T_g$, it is left to be 
investigated how $\cot \theta_H(T)$ behaves around $T_g$.

Since the pseudogap effect is expected to be related to the
antiferromagnetic fluctuations \cite{Batlogg}, there have been efforts 
to investigate how the pseudogap feature is affected by 
Zn doping onto the CuO$_2$ planes, which produces spin vacancies.
The reported Zn-doping effects on the pseudogap are not simple;
for example, the pseudogap feature in 
$\rho_{ab}(T)$ in underdoped YBCO crystals is almost unchanged
\cite{Mizuhashi},
while the suppression in the Cu NMR relaxation rate below $T_g$ is 
diminished with only 1\% of Zn \cite{Zheng}.
To build a complete picture of the pseudogap effect, it is also 
useful to investigate how the Zn doping affects the pseudogap in the 
Hall channel.

In this paper, we report the results of our measurements
of the Hall effect in YBa$_{2}$(Cu$_{1-z}$Zn$_z$)$_{3}$O$_{y}$ 
crystals with $y$=6.78, which corresponds to an underdoped concentration.
At this composition $y$=6.78, 
which gives $T_c$$\simeq$75 K in pure crystals, 
a peak in $R_H(T)$ can be clearly seen 
and also the pseudogap feature in $\rho_{ab}(T)$ is clearly discernible 
(due to the rather wide $T$-linear region above $T^*$); from the
literature, we can infer that $T^*$ is about 200 K (Ref. \cite{Ito}) 
and $T_g$ is about 130 K (Ref. \cite{Diagram}).
Our measurements of three samples with different Zn concentrations 
(z=0, 0.006, and 0.013) found that a deviation from the $T^2$ 
behavior in $\cot \theta_H$ takes place in all the samples at the same 
temperature $T_0$ which is very close to $T_g$, indicating that the pseudogap indeed affects $\cot \theta_H$ near $T_g$ and that the effect 
is robust against Zn doping.

There have been several publications reporting the effect of Zn doping
on $R_H$ in YBCO, but the results are not converged. 
The data by Chien, Wang, and Ong indicate 
that $R_H$ of optimally-doped
crystals increases with increasing $z$ in the whole
temperature range above $T_c$ and the $T$ dependence becomes less 
pronounced \cite{Chien} [it is possible that
in their samples the effective carrier concentration is changing,
because the slope of $\rho_{ab}(T)$ is increasing with $z$].
Mizuhashi {\it et al.} reported  
that $R_H$ increases over the whole
temperature range with $z$ (almost like a parallel shifting), 
while the slope of $\rho_{ab}(T)$ in the $T$-linear part is unchanged 
\cite{Mizuhashi}.
On the other hand, Walker, Mackenzie, and Cooper reported 
that, in their Zn-doped crystalline thin films, $R_H$ 
at 300 K remains essentially unchanged, while at low temperatures $R_H$ 
is progressively suppressed with increasing $z$ \cite{Cooper,Walker}.  
In the present work, 
we therefore paid particular attention to reduce the errors in the 
measurement of $R_H$; the Hall voltage is measured with 
magnetic-field sweeps at constant temperatures, 
and errors due to the geometrical factors 
are minimized by making small voltage contacts and by determining 
the sample thickness with a high accuracy. 
We note that making the voltage contacts on the side faces (not on the 
top face) of the crystals is essential in reducing the error and
increasing the reproducibility.

\begin{figure}
\epsfxsize=0.8\columnwidth
\centerline{\epsffile{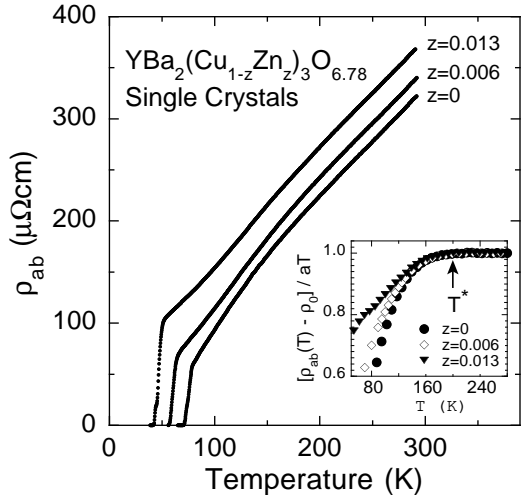}}
\vspace{0.2cm}
\caption{
$T$ dependence of $\rho_{ab}$ for the pure and Zn-doped samples.
Inset: Plots of $(\rho_{ab}(T)-\rho_0)/aT$ vs $T$, where  
$\rho_0$=13.9, 34.6, and 63.4 ${\rm \mu\Omega}$cm 
for $z$=0, 0.006, and 0.013, 
respectively.  The slope $a$ (=1.05) is unchanged with $z$.
$T^*$ is marked by an arrow.}
\label{fig1}
\end{figure}

The Zn-doped YBCO single crystals are grown by a flux method using pure 
Y$_2$O$_3$ crucibles \cite{Segawa}.  All the crystals measured here are
naturally twinned.  The oxygen content is tuned to $y$=6.78 by 
annealing the crystals with pure YBCO powders in air at 575$^{\circ}$C 
for 37 h, and subsequent quenching to room temperature.
The final oxygen content is confirmed by iodometric titration.
The actual Zn concentration in the crystals are measured with
the inductively-coupled plasma (ICP) spectrometry with an error in $z$ 
of less than $\pm$0.001.

\begin{figure}
\epsfxsize=0.8\columnwidth
\centerline{\epsffile{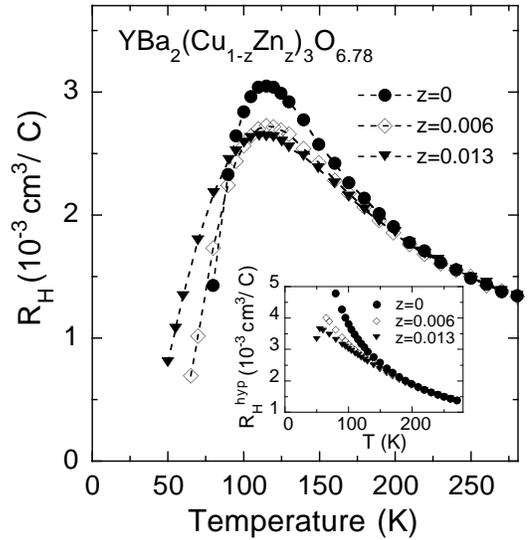}}
\vspace{0.2cm}
\caption{
$T$ dependence of $R_H$ for pure and Zn-doped samples.
Inset: Plot of $R_H^{hyp}$ vs $T$ for the three samples, see text.}
\label{fig2}
\end{figure}

The measurements are performed with a low-frequency (16 Hz) ac
technique.  Longitudinal and transverse voltages are measured 
simultaneously using two lock-in amplifiers during the field sweeps at
constant temperatures.  For the transverse signal, we achieved a high
sensitivity by subtracting the offset voltage at zero field 
(the offset comes from a slight longitudinal misalignment between the 
two Hall voltage contacts).
The temperature is stabilized using a high-resolution resistance
bridge with a Cernox resistance thermometer.  We confined the maximum 
magnetic field to 4 T, with which the error of the Cernox thermometer 
caused by its own magnetoresistance is negligibly small in the 
temperature range of the present study. 
The magnetic field is applied along the $c$-axis of the crystals.
To enhance the temperature stability, the sample and the thermometer are 
placed in a vacuum can with a weak thermal link to the outside.
The achieved stability in temperature during the field sweeps is 
better than a few mK.
The data are taken from $-4$ T to $+4$ T, and then the 
asymmetrical component
is calculated to obtain the true Hall voltage.
The final accuracy in the magnitude of $R_H$ and $\rho_{ab}$ reported here 
is estimated to be better than $\pm$5\%, and the relative error in the
data for each sample is less than $\pm$2\%.

Figure 1 shows the temperature dependence of $\rho_{ab}$ for the
three Zn concentrations.  
Above $\sim$200 K, $\rho_{ab}$ of all the three samples shows  
a good $T$-linear behavior and the slope of this $T$-linear part 
does not change with $z$.  As shown in the inset to Fig. 1, a 
downward deviation from the $T$-linear dependence takes place at
the same temperature for all the three samples, indicating that the
upper pseudogap temperature $T^*$ does not change with $z$.  
This result is in good agreement with the previous reports 
\cite{Mizuhashi,Walker}.

Figure 2 shows the temperature dependence of $R_H$ for the three samples.
Our results are somewhat different from previous results on single crystals
\cite{Chien,Mizuhashi}, but rather resemble that of the thin film result
\cite{Walker}.  Notably, $R_H$ around 250 K does not change with 
$z$, while the peak at 110 K is clearly suppressed with increasing
Zn concentration.
Still, the behavior of $\cot \theta_H$ is in good agreement with the
previous studies; as is shown in Fig. 3, $\cot \theta_H$ 
changes approximately as $T^2$ in a rather wide range, and the Zn impurities
add a $T$-independent offset which is roughly proportional to $z$.

\begin{figure}
\epsfxsize=0.7\columnwidth
\vspace{0.3cm}
\centerline{\epsffile{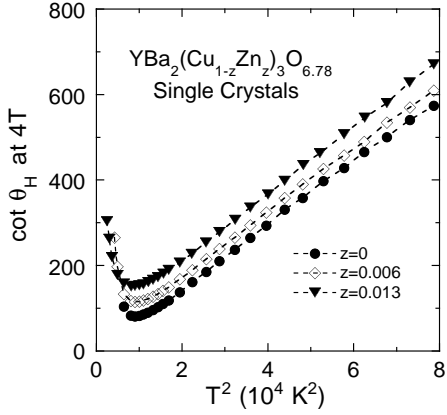}}
\caption{
Plots of $\cot \theta_H$ vs $T^2$ for the three samples.}
\label{fig3}
\end{figure}

We note that the Zn-doping effect on $R_H(T)$ observed here is 
naturally expected in the context of the two scattering time scenario.
One can infer that the primary effect of Zn-doping is to add some
constant impurity-scattering rates to both $\tau_{tr}^{-1}$ and 
$\tau_H^{-1}$, because both $\rho_{ab}(T)$ and $\cot \theta_H(T)$ show 
essentially parallel shifts upon Zn-doping.
Since one can approximately express 
$\tau_{tr}^{-1}$$\sim$$T$ and $\tau_H^{-1}$$\sim$$T^2$ in pure samples, 
the scattering rates in Zn-doped samples can be approximated as
$\tau_{tr}^{-1} \sim T+A$ and $\tau_H^{-1} \sim T^2+B$.
From the relation $R_H H$ = $\rho_{ab}/\cot \theta_H$ 
$\sim \tau_H/\tau_{tr}$,
$R_H$ is approximately written as $R_H \sim (T+A)/(T^2+B)$ 
in Zn-doped samples.  
If we compare this expression with that for the pure samples,
$R_H^{pure} \sim T/T^2 \sim T^{-1}$, we can infer that 
at high temperatures
$R_H$ in Zn-doped sample should approach $R_H^{pure}$, 
while at low temperatures $R_H$ in Zn-doped 
sample is expected to become smaller than $R_H^{pure}$ 
(which can be easily seen when one considers $T$$\to$0).
The above heuristic argument implies that the weakening of the 
$T$ dependence of $R_H(T)$, combined with a $z$-independent 
room-temperature $R_H$, is a rather natural consequence of the 
Zn-doping in the two scattering time scenario, 
although this effect has not been well documented before.

\begin{figure}
\epsfxsize=0.7\columnwidth
\centerline{\epsffile{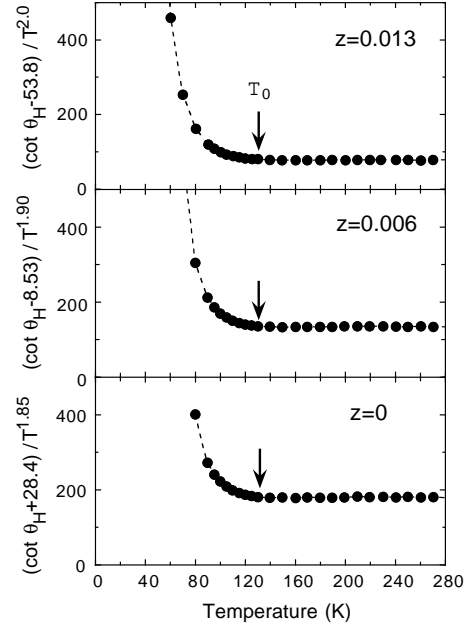}}
\vspace{0.2cm}
\caption{
Plots of $(\cot \theta_H - C)/T^{\alpha}$ vs $T$,
which emphasizes where the the deviation from 
the high-temperature behavior $\cot \theta_H$ = $C+DT^{\alpha}$ 
(with $\alpha$$\simeq$2) takes place. 
The deviation at $T_0$ is marked by arrows.}
\label{fig4}
\end{figure}

Now let us analyze the data in more detail in regard of the $T$ 
dependence of $\cot \theta_H$.
A close examination of Fig. 3 tells us that the data for $z$=0 and 0.006 
are slightly curved in this plot; we found that the best power laws 
to describe the data in a wide temperature range are 
$T^{1.85}$, $T^{1.9}$, and $T^{2.0}$, for
$z$=0, 0.006, and 0.013, respectively.
In Fig. 4, we show plots of $(\cot \theta_H - C)/T^{\alpha}$ vs $T$,
which cancels out the power-law temperature dependence and therefore 
we can easily see the temperature range for the $T^{\alpha}$ dependence 
to hold well.
Here, $C$ is the offset value (which increases with $z$) 
and $\alpha$ is the best power for each Zn concentration.
It is clear from Fig. 4 that the power-law temperature dependence of 
$\cot \theta_H$ holds very well down to a temperature $T_0$ ($\sim$130 K)
and then starts to deviate in all the three samples.
Incidentally, the deviation occurs at a temperature very close to
$T_g$, which is $\sim$130 K for $y$$\approx$6.78 (Ref. \cite{Diagram}). 
This is a strong indication that the change in $\cot \theta_H(T)$ is 
caused by the opening of the pseudogap \cite{Note}.
Our result shows that, unlike the Cu NMR relaxation rate, 
the Zn-doping does not diminish or shift the onset of the pseudogap 
marked by the change in $\cot \theta_H$ at $T_0$, at least up to the
Zn concentration of 1.3\%.
Note, however, that the deviation from the power law becomes a bit
weaker (or slower) with increasing $z$, which is similar to what is 
seen in the behavior of $\rho_{ab}(T)$ (inset to Fig. 1).

Given the fact that $\cot \theta_H$ is apparently affected by the pseudogap
below $T_0$, it is useful to clarify how the pseudogap effect is reflected 
in the $T$ dependence of $R_H$, 
which is a result of the two different $T$ dependences \cite{Chien} 
of the more fundamental parameters $\tau_{tr}^{-1}$ and $\tau_H^{-1}$.
For this purpose, it is instructive to see how $R_H(T)$ would behave 
if $\cot \theta_H$ continues to change as $T^{\alpha}$ down to $T_c$.
The inset to Fig. 2 shows the plots 
of the $T$ dependence of such hypothetical $R_H^{hyp}$ for the three 
samples, where $R_H^{hyp}$ is calculated by dividing $\rho_{ab}$ by 
$(C+DT^{\alpha})\times H$, where $D$ is the $T$-independent value 
at temperatures above $T_0$ in Fig. 3.  
It is clear from the behavior of $R_H^{hyp}$ 
that $R_H(T)$ would {\it not} show a peak if $\cot \theta_H$ 
continues to change as $T^{\alpha}$ down to $T_c$.
Therefore, we can conclude that the peak in $R_H(T)$ in underdoped
YBCO is caused by the opening of the pseudogap.

It should be noted that the direction of the change in $\cot \theta_H$ 
at $T_0$ implies that $\tau_H^{-1}$ is {\it enhanced} when the 
pseudogap opens; this is opposite to the effect on $\tau_{tr}^{-1}$, 
which is {\it reduced} below $T^*$.  Therefore, we cannot
simply conclude that the change in $\cot \theta_H$ is caused by a
reduced electron-electron scattering, which is the natural consequence 
of a pseudogap in the low-energy electronic excitations.  
One possibility to understand this apparently 
confusing fact is to attribute the change at $T_0$ to the effective mass, 
rather than to attribute it to the scattering rate; remember that 
$\cot \theta_H$ = $1/(\omega_c \tau_H)$ $\propto m_H/\tau_H$, 
where $m_H$ is the effective mass of the particle responsible for the
Hall channel \cite{Chien}, 
so an increase in $\cot \theta_H$ is expected when the 
effective mass is enhanced.  For example, if the pseudogap is related to the
formation of a dynamical charged stripes \cite{Emery}, a modification
in the FS topology, which leads to a change in the
effective mass, is expected.
This picture is also consistent with the observed robustness of the 
pseudogap feature in 
$\cot \theta_H$ upon Zn doping, because the change in the FS topology
is rather insensitive to a small amount of impurities.
One might question why there is little trace of the effective-mass change
in the $T$ dependence of $\rho_{ab}$. 
If $\cot \theta_H$ and $\rho_{ab}$ reflect different parts of the FS 
(as is conjectured in the hot/cold spots scenario \cite{Hot/Cold}), 
it is possible that the modification of the FS topology alters 
the band mass for the Hall channel while leaving that of the diagonal
channel relatively unchanged.

Finally, we note that the peak in the $T$ dependence of $R_H$ 
is not always caused by the pseudogap.  For example, 
in overdoped Tl$_2$Ba$_2$CuO$_{6+\delta}$ (Tl-2201), 
it has been reported \cite{Kubo} that
$\cot \theta_H$ shows a good $T^2$ dependence down to near $T_c$ 
(which implies that the pseudogap does not open),
and yet the peak in $R_H(T)$ is observed at a temperature 
well above $T_c$.
In this case, the peak in $R_H(T)$ is just a result of the two
different $T$ dependences of $\tau_{tr}^{-1}\sim T^n+A$ (1$\le$$n$$\le$1.9) 
and $\tau_H^{-1}\sim T^2+B$ (note that in Tl-2201 both 
$\tau_{tr}^{-1}$ and $\tau_H^{-1}$ have somewhat large offsets 
even in pure crystals \cite{Kubo}).  Mathematically, 
$R_H \sim (T^n+A)/(T^2+B)$ has a peaked $T$-dependence and thus 
$R_H(T)$ can show a peak well above $T_c$ for some combination of 
$A$ and $B$, even when both $\rho_{ab}$ and $\cot \theta_H$ 
do not show any deviation from the power laws.
On the other hand, as is demonstrated in the inset to Fig. 2, 
the peak in $R_H(T)$ of underdoped YBCO cannot be accounted for by the 
above origin and therefore is clearly caused by the pseudogap.
This argument tells us that one should always look at the $T$ dependence
of $\cot \theta_H$, not just the peak in $R_H(T)$, to determine whether
the pseudogap is showing up through $(\omega_c \tau_H)^{-1}$.

In summary, we observed that $\cot \theta_H$ of pure and Zn-doped 
YBCO ($y$=6.78) crystals shows an upward deviation 
from the $T^2$ behavior below a temperature $T_0$ that is notably 
higher than $T_c$ but is much lower than $T^*$.  
The onset temperature $T_0$ for this 
deviation, which is found to be unaffected by Zn doping, 
is close to the lower temperature scale for the pseudogap
$T_g$ (probed by the Cu NMR relaxation rate, for example).  
The fact that 
$\cot \theta_H$ tends to be {\it enhanced} below $T_0$ suggests that the
effect of the pseudogap is {\it not} to reduce the Hall scattering rate;
we therefore propose that the effect is more likely to be originating
from a change in the Fermi surface topology, which causes a change 
in the effective mass.
Also, we demonstrated that the peak in $R_H(T)$ of underdoped YBCO is 
not just a result of two different scattering times, but is actually 
a result of the pseudogap effect on $\cot \theta_H$.

We thank A. N. Lavrov and I. Tsukada for fruitful discussions, and 
J. Takeya for technical assistance.

%
\medskip
\vfil

\end{document}